\documentstyle[epsf]{cupconf}

\ifoldfss
\else
  \ifnfssone
    \newmathalphabet{\mathit}
      \addtoversion{normal}{\mathit}{cmr}{m}{it}
      \addtoversion{bold}{\mathit}{cmr}{bx}{it}
    \newmathalphabet{\mathcal}
      \addtoversion{normal}{\mathcal}{cmsy}{m}{n}
    \else
    \ifnfsstwo
    \fi
  \fi
\fi

\newcommand{\Msol}{\mbox{$M_{\odot}$}}

\newcommand{\fcon}{\mbox{$f_{\rm con}$}}
\newcommand{\Xco}{\mbox{$X_{\rm CO}$}}
\def\hexnumber#1{\ifcase#1 0\or1\or2\or3\or4\or5\or6\or7\or8\or9\or
 A\or B\or C\or D\or E\or F\fi }

\makeatletter
\ifx\CUP@mtlplain@loaded\undefined
\else
\fi
\makeatother
 \makeatletter
 \ifx\CUP@mtlplain@loaded\undefined
   \font\tenbmi=cmmib10 at 10pt
   \font\sevenbmi=cmmib10 at 7pt
   \font\fivebmi=cmmib10 at 5pt

   \newfam\bmifam
   \textfont\bmifam=\tenbmi
   \scriptfont\bmifam=\sevenbmi
   \scriptscriptfont\bmifam=\fivebmi
   
 \fi
 \makeatother
\ifnfsstwo

\fi
\ifnfssone

\fi
\ifoldfss

\fi

\mathchardef\varLambda="0103

\makeatletter
\ifx\CUP@mtlplain@loaded\undefined
\else
\fi
\makeatother
\makeatletter
\ifx\CUP@mtlplain@loaded\undefined
  \font\tenbms=cmbsy10
  \font\sevenbms=cmbsy10 at 7pt
  \font\fivebms=cmbsy10 at 5pt
  \newfam\bmsfam
  \textfont\bmsfam=\tenbms
  \scriptfont\bmsfam=\sevenbms
  \scriptscriptfont\bmsfam=\fivebms

  \edef\bsy@{\hexnumber\bmsfam}
  \mathchardef\bnabla="0\bsy@72
\fi
\makeatother
\def\etal{\mbox{\it et al.}}

\title[Radial Transport of Molecular Gas]{Radial Transport of Molecular Gas
to the Nuclei of Spiral Galaxies}

\author[K. Sakamoto {\it et al.\/}]%
{K.\ns  S\ls A\ls K\ls A\ls M\ls O\ls T\ls O$^{1,2}$,\ns
 S.\ns  K.\ns  O\ls K\ls U\ls M\ls U\ls R\ls A$^{1}$\ls,\ns \\
 S.\ns  I\ls S\ls H\ls I\ls Z\ls U\ls K\ls I$^{1}$,\ns 
 and \ns
 N.\ns  Z.\ns     S\ls C\ls O\ls V\ls I\ls L\ls L\ls E$^{2}$}

\affiliation{$^1$Nobeyama Radio Observatory, Nagano, 384-1305, JAPAN\\[\affilskip]
$^2$Radio Astronomy, California Institute of Technology, MS105-24,
Pasadena, CA91125, USA}

\begin{document}
\ifnfssone
\else
  \ifnfsstwo
  \else
    \ifoldfss
      \let\mathcal\cal
      \let\mathrm\rm
      \let\mathsf\sf
    \fi
  \fi
\fi

\maketitle

\begin{abstract}
The NRO/OVRO imaging survey of molecular gas in 20 spiral galaxies
is used to test the theoretical predictions on bar-driven gas transport,
bar dissolution, and bulge evolution.
In most galaxies in the survey we find gas condensations
of 10$^8$--10$^9$ \Msol\ within the central kiloparsec,
the gas masses being comparable to those needed to 
destroy bars in numerical models. 
We also find a statistically significant difference in the degree 
of gas concentration between barred and unbarred galaxies: 
molecular gas is more concentrated to the central kiloparsec in barred systems. 
The latter result supports the theories of bar-driven gas transport. 
Moreover, it constrains the balance between
the rate of gas inflow and that of gas consumption (i.e., star formation, etc.),
and also constrains the timescale of the possible bar dissolution.
Namely, gas inflow rates to the central kiloparsec, averaged over the 
ages of the bars, must be larger than the mean rates of 
gas consumption in the central regions in order
to cause and maintain the higher
gas concentrations in barred galaxies.
Also, the timescale for bar dissolution must be longer 
than that for gas consumption in the central regions
by the same token.
\end{abstract}

\firstsection % if your document starts with a section,
\section{Introduction}
	Radial transport of gas in galactic disks likely plays an important
role in the formation and evolution of bulges.
There are two aspects in the effect of gas transfer to bulges, in
both of which stellar bars are involved.
First, theories predict that bars efficiently
transport interstellar gas to the nuclei of spiral galaxies, 
providing star forming material to the bulge regions.
Second, simulations have shown that 
the gas accumulation at a galactic center changes the
gravitational potential and eventually destroys the bar 
(c.f., a review by Pfenniger in this workshop).
Bulges may grow through this process by gaining stars from disks.

	Observationally, evidence for bar-driven gas transport
and for bar dissolution
has been limited compared to the large amount of theoretical works.
The pieces of observational evidence supporting 
the bar-driven gas transport are
the estimation of gas inflow rates in two barred galaxies using
CO and NIR observations and dynamical models 
(Quillen \etal\ 1995; Regan \& Vogel 1997), 
shallower metallicity gradients in barred than unbarred galaxies 
(Zaritsky \etal\ 1994; Martin \& Roy 1994), and larger
H$\alpha$ luminosities in the nuclei of barred galaxies 
presumably due to larger amount of gas in barred nuclei (e.g., Ho \etal\ 1995).
In order to further investigate the relation between bars, gas, and
bulges, it is important to observe gas in many galaxies.

	The NRO/OVRO CO imaging survey mapped the distribution of molecular
gas in the central kiloparsecs of 20 ordinary nearby spirals 
using the millimeter arrays of the two observatories
(Sakamoto \etal\ 1998, 1999). 
The 20 northern spiral galaxies were
selected on the basis of inclination (face-on),
lack of significant dynamical perturbation, and reasonable single-dish
CO flux to allow high-resolution observations.
No selection was made on starburst, nuclear activity, far-infrared
luminosity, and galaxy morphologies.
The sample contains 10 barred (SB+SAB) and 10 unbarred (SA) spirals
with the mean distance of 15 Mpc and with luminosities $\sim L^{*}$.
Our aperture synthesis observations have a mean resolution of $4''$ 
(= 300 pc at 15 Mpc) 
and recovered most ($70 \pm 14$ \%) of the single-dish flux.
We use the data to set constraints on the above theoretical predictions.

\section{Central gas condensations}
	Most galaxies in our sample show strong condensations of CO 
at their centers.
Fig. 1 shows the histogram of CO-derived masses of molecular gas 
within the central kiloparsec. 
The central gas masses are mostly in the range of 10$^8$--10$^9$ \Msol.
It thus seems not unusual for a large gas-rich galaxy to have a 
condensation of such a large amount of gas at the center.
The gas condensations generally have radial profiles sharply peaking toward
the galactic centers, when observed with sub-kiloparsec resolutions.
The distribution of radial scale lengths of CO is also in Fig. 1. 
The central scale length is defined as the radius at which a radial profile 
falls to $1/e$ of its maximum value, and is not affected much by
the missing flux (15 \% error at most).
It is apparent that most galaxies have sub-kiloparsec scale lengths
in the nuclear regions.
The gas condensations are thus not simple extensions of 
outer exponential disks, which usually have scale lengths larger than  
a few kpc.
It is interesting to note that the highest mass of the gas condensations, 
10$^9$ \Msol, is comparable to the mass needed to destroy bars in
simulations.

\begin{figure} 
%  \vspace{16.5pc}
{\hfill
\epsfysize=4.8cm\epsfbox{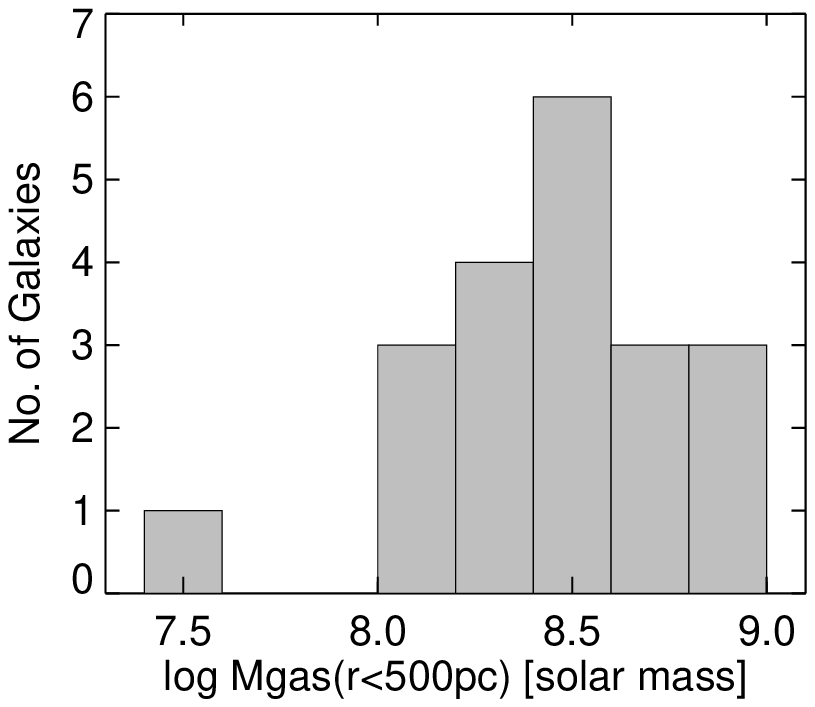}
\hspace{0.5cm}
\epsfysize=4.8cm\epsfbox{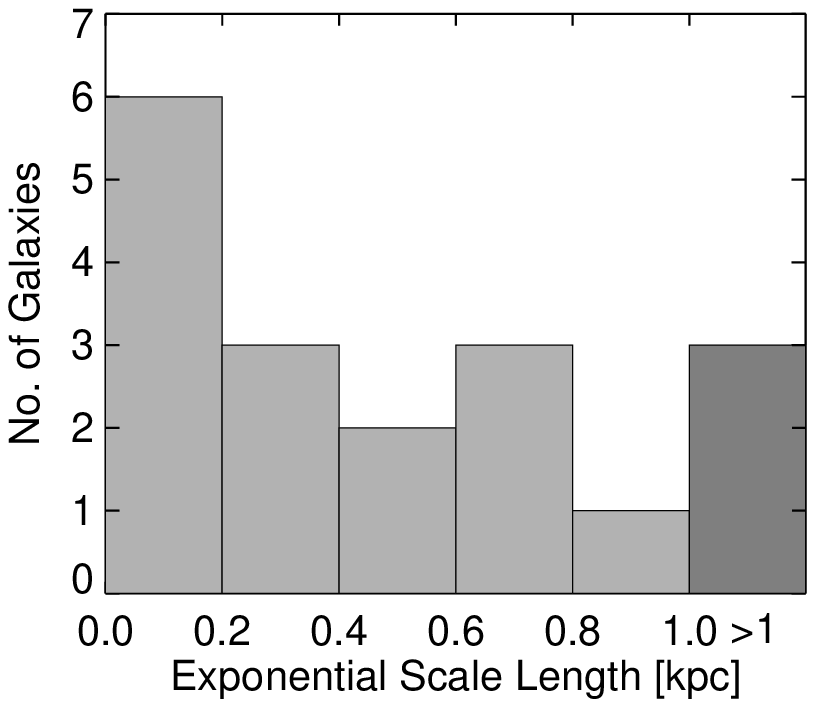}
\hfill}
  \caption{(Left) Molecular gas masses within the central kiloparsec derived
from CO emission. 
(Right) Scale lengths of CO radial distributions in the galactic centers.} 
\end{figure} 

\section{Higher gas concentrations in barred galaxies}
	In order to quantify the degree of gas concentration
in disk galaxies, we compare in Fig. 2 the gas surface densities
averaged in the central kiloparsec with those averaged over the optical
galactic disks (i.e., $R<R_{25}$). 
The former are calculated from our data and 
the latter are calculated from the single-dish mapping data of the FCRAO
survey (Young \etal\ 1995).
The ratio of the two surface densities, 
$f_{\rm con} \equiv 
\Sigma_{\rm gas}^{R<500 \rm pc}/\Sigma_{\rm gas}^{R<R_{25}}$, 
is an indicator of gas concentration to the central kiloparsec.
The concentration factor \fcon\ shows more than 100-fold variation in our
sample.

	Barred (i.e., SB+SAB) and unbarred (SA) galaxies are
plotted with different symbols in Fig. 2. 
It is apparent that barred galaxies 
have higher concentration factors than unbarred galaxies.
The difference is statistically significant according to the
Kolmogorov-Smirnov test; the probability to observe the
difference of \fcon\ in Fig. 2 would be 0.007 
if there were no difference between the two classes of galaxies.

\begin{figure} 
%  \vspace{16.5pc}
{\hfill
\epsfysize=7cm\epsfbox{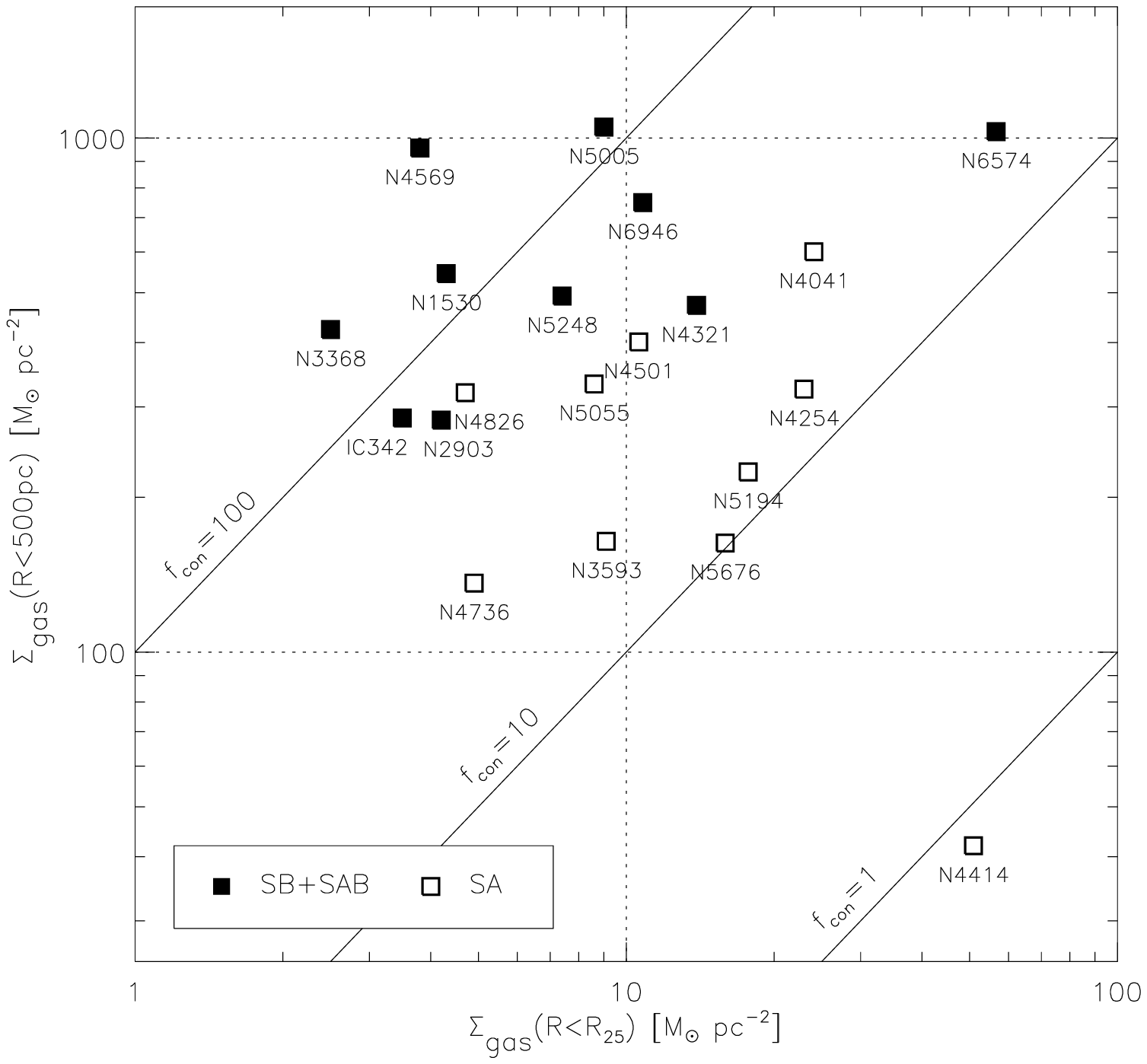}
\hspace{0.5cm}
\epsfysize=7cm\epsfbox{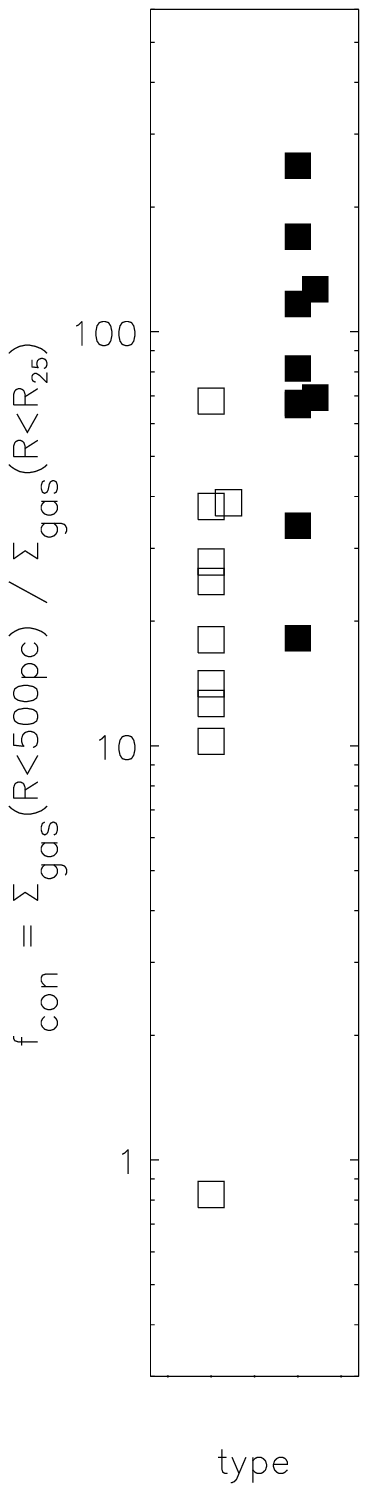}
\hfill}
  \caption{(Left) Surface densities of molecular gas averaged within the central
kiloparsec are compared to those averaged over the optical galactic disks.
The ratio of the central-to-disk averaged surface densities is an
index of gas central concentration.
Galaxies in the upper-left part of the panel have higher ratios, i.e.,
higher gas concentrations.
(Right) Distributions of the surface density ratios (i.e., concentration
factors \fcon) for barred and unbarred galaxies.} 
\end{figure} 

	We note that the conversion factor from CO flux to mass of 
molecular gas, $X_{\rm CO}$, is unlikely to produce the difference
of \fcon\ 
between barred and unbarred galaxies.
The ratio $f_{\rm con}$ is independent of \Xco\ 
if it has the same form of radial distribution in galaxies, 
e.g., $\Xco(r)$ being either $a$ (const.), $ae^{br}$, or $ar^{b}$
with the same radial scale $b$.
The multiplier $a$ can be
different from one galaxy to another without changing $f_{\rm con}$.
A systematic difference in the radial 
profile of \Xco\ between barred and unbarred galaxies
may exist if the conversion factor scales with metallicity.
However, the shallower metallicity gradients in barred galaxies
would make the apparent CO concentrations lower in barred galaxies.
Thus the correction for metallicity would only enhance the difference of
$f_{\rm con}$ between the two types of galaxies.
No other cause is known to create a systematic difference in
radial profiles of \Xco\ between barred and unbarred systems.

	We used classifications in the Third Reference Catalogue
to distinguish barred and unbarred galaxies. 
It is possible that the classifications based on
optical images missed small nuclear bars or misidentified open spiral
arms as a bar. 
However, nuclear bars naturally have smaller power of gas transport,
and spiral arms masquerading a bar create a global nonaxisymmetry in 
the gravitational potential as a bar does.
Thus the optical classification is a qualitative index of
the strength of nonaxisymmetry in mass distribution and in gravitational
potential.
We conclude therefore that galaxies with
larger nonaxisymmetries (called `barred' galaxies) have 
higher gas concentrations
than galaxies with smaller nonaxisymmetries (i.e., `unbarred' galaxies)
\footnote[1]{The gas concentrations in {\it unbarred} galaxies 
are low but $\geq 1$.
They may be due to the bars that had been destroyed, but they do not 
necessarily require the bar dissolution, because
they may be also due to viscous accretion of gas, 
weak but finite nonaxisymmetries in those galaxies,
or the centrally peaked distribution of stars that produce gas.
}.

\section{Implications to the bar-dissolution scenario}
	The higher gas concentrations in barred galaxies are
most likely due to radial transport of gas in the barred potentials.
However, the transport of gas is not a sufficient condition
to cause and maintain the higher gas concentrations in barred galaxies.
It is also necessary that the molecular gas funneled to the
galactic centers remains there in molecular form {\it and} that
the bars responsible for the gas transport remain.
These requirements set constraints on the relation between the rates
of gas inflow and gas consumption,
and also on the timescale for the possible bar dissolution.

	First, the total amount of gas funneled to the center of a barred
galaxy must be larger than the total amount of stars formed in the
same region, because otherwise the higher gas concentration in the barred
galaxy can not be sustained. 
Dividing the total masses by the age of the bar, 
the above relation translates to the condition
that the time-averaged rate of gas inflow must be
larger than that of star formation.
One may be able to estimate the time-averaged rate of star formation 
from an ensemble-average of star formation rates in the centers of 
barred galaxies, thereby setting a lower limit to the mean gas inflow rate.
If there are other ways of loosing molecular gas, such as a
gas outflow due to starburst and accretion to active nucleus, then
the lower limit becomes higher.

	The second condition we can deduce is that the timescale of
gas consumption in the central regions is longer than that of the 
possible bar dissolution. 
In other words, if bars are to be destroyed by the gas inflow
of 10$^8$--10$^9$ \Msol\ to the central kiloparsec and if the bar
dissolution is much quicker than the gas consumption in the
central regions, then we would see currently unbarred but previously
barred galaxies with high central gas concentrations that destroyed
the bars. 
The lack of such galaxies (i.e., unbarred spirals with
$f_{\rm con} \geq 100$) allows us to set the above condition on the
timescale of bar dissolution.

	Quantitative evaluation of the above conditions is hampered
by the difficulty in accurately estimating star formation rates 
in galactic centers. 
The current star formation rates crudely estimated from
H$\alpha$ in the centers of the sample galaxies are 
$\sim 0.1$--1 \Msol yr$^{-1}$, 
which set a lower limit to the mass inflow rate.
The consumption time of the gas concentrations is 10$^8$--10$^{10}$ yrs. 
The lower value does not contradict with the predicted timescale of
bar dissolution, which is comparable to the dynamical time or a few 10$^8$ yrs.
If the higher value is the case in many spirals, then the bar dissolution
must take longer time than predicted, or will not happen
for the 10$^8$--10$^9$ \Msol\ gas concentrations.

	It seems worthwhile to compile more data of gas concentration
and star formation to tighten the above constraints on the mass transfer
in galactic disks and on the fate of stellar bars.
The index of gas concentration \fcon\ may be useable as a tool to
find out unbarred galaxies that were barred and galaxies with young bars: 
the former must have higher concentration factors for unbarred galaxies 
and the latter must have lower factors for barred galaxies.
Observations of such galaxies would tell us about evolution of disks,
bars, and bulges.

\begin{acknowledgments}
Stimulating conversations at the workshop with Drs. Norman, Pfenniger, Hasan, Regan, and Wada are acknowledged.
K.S. was supported by JSPS grant-in-aid.
\end{acknowledgments}

\end{document}